\documentclass[aps,prl,twocolumn,superscriptaddress,amsfont,graphicx,nofootinbib,preprintnumbers]{revtex4}%
\usepackage{color,graphicx,epsfig}
\usepackage{ifpdf}
\usepackage{amsmath}
\usepackage{bm}
\usepackage{color}
\usepackage[english]{babel}
\usepackage{graphicx}%
\usepackage{amsfonts}%
\usepackage{amssymb}
\usepackage{braket}
\usepackage{hyperref}
\usepackage{enumerate}

\definecolor{nicered}{rgb}{0.7,0.1,0.1}
\definecolor{nicegreen}{rgb}{0.1,0.5,0.1}
\hypersetup{colorlinks,citecolor= nicegreen,linkcolor= nicered}

\usepackage{graphicx}
\usepackage{dcolumn}
\usepackage{bm}
\usepackage{slashed}

\begin{document}

\title{Leptophilic dark matter in gauged $U(1)_{L_e-L_\mu}$ model in light of DAMPE cosmic ray $e^+ + e^-$ excess}

\author{Guang Hua Duan}
\email[Electronic address:]{ghduan@itp.ac.cn}
\affiliation{Department of Physics and Institute of Theoretical Physics, \\Nanjing Normal University, Nanjing, Jiangsu 210023, China}
\affiliation{CAS Key Laboratory of Theoretical Physics, Institute of Theoretical Physics, \\ Academia Sinica, Beijing 100190, China}
\affiliation{School of Physical Sciences, \\ University of Chinese Academy of Sciences, Beijing 100049, China}

\author{Xiao-Gang He}
\email[Electronic address:]{hexg@sjtu.edu.cn}
\affiliation{Department of Physics, National Taiwan University, Taipei 10617}
\affiliation{Physics Division, National Center for Theoretical Sciences, Hsinchu 30013}
\affiliation{T-D Lee Institute, School of Physics and Astronomy, \\ Shanghai Jiao Tong University, Shanghai 200240, China}

\author{Lei Wu}
\email[Electronic address:]{leiwu@itp.ac.cn}
\affiliation{Department of Physics and Institute of Theoretical Physics, \\Nanjing Normal University, Nanjing, Jiangsu 210023, China}

\author{Jin Min Yang}
\email[Electronic address:]{jmyang@itp.ac.cn}
\affiliation{CAS Key Laboratory of Theoretical Physics, Institute of Theoretical Physics, \\ Academia Sinica, Beijing 100190, China}
\affiliation{School of Physical Sciences, \\ University of Chinese Academy of Sciences, Beijing 100049, China}
\affiliation{Department of Physics, Tohoku University, Sendai 980-8578, Japan}

\date{\today}

\begin{abstract}
Motivated by the very recent cosmic-ray electron+positron excess observed by DAMPE collaboration, we investigate a Dirac fermion dark matter (DM) in the gauged $L_e - L_\mu$ model. DM interacts with the electron and muon via the $U(1)_{e-\mu}$ gauge boson $Z^\prime$. The model
can explain the DAMPE data well. Although a non-zero DM-nucleon cross section is only generated at one loop level and there is a partial cancellation between $Z^{\prime}ee$ and $Z^{\prime}\mu\mu$ couplings, we find that a large portion of $Z\prime$ mass is ruled out from direct DM detection limit leaving the allowed $Z^{\prime}$ mass to be close to two times of the DM mass.
Implications for $pp \to Z^{\prime} \to 2\ell$ and $pp \to 2\ell + Z^{\prime}$, and muon $g-2$ anomaly are also studied.
\end{abstract}
\pacs{Valid PACS appear here}
\maketitle


\section{Introduction}
Many cosmological observations have established a standard cosmological model, in which the Dark Matter accounts for about 27\% of the global energy budget. Cold Dark Matter (CDM) provides a natural way to produce theoretical predictions in striking agreement with observations. On the other hand, the nature of CDM is still unknown.
The paradigm of weakly interacting massive particles (WIMPs) is one of the most compelling versions.

Up to now, the WIMP dark matter has been scrutinized in various underground and collider experiments. The recent limits of measuring nucleon-WIMP dark matter scattering have already excluded a large portion of WIMP parameter space~\cite{pandaX-II,xenon-1t}, approaching the neutrino floor. Besides, the null results from LHC searches for mono-X signatures also put stringent constraints on WIMP DM candidates~\cite{dm-review}. However, the DM indirect detections, such as AMS-02, PAMELA, HEAT and Fermi-LAT, have reported some intriguing evidences of DM, which inclines to annihilate into leptons. No excess in the anti-proton flux has been observed. Given these observations, the DM may have no interactions with quarks at tree level. Furthermore, the very recent results of measuring cosmic-ray electrons and positions by DAMPE collaboration~\cite{dampe-mission} has shown a sharp peak at $\sim 1.4 $ TeV in $e^+ + e^-$ flux~\cite{dampe-data}.
In Ref.~\cite{yuan2017,bi}, the authors analyzed the data and discussed the astrophysical and DM interpretations. One possible way to explain the data is that the DMs annihilate into leptons and the mass of DM particle is about 1.5 TeV if the nearby DM sub-halo located at $\rm 0.1 \sim 0.3$ kpc away from solar system ~\cite{yuan2017}. Several leptophilic DM models have been proposed to explain this excess~\cite{Fan:2017sor,Gu:2017gle,Duan:2017pkq}.

In this work, we explain this tentative DAMPE $e^+ + e^-$ excess in a gauged $L_e-L_\mu$ model with a Dirac fermion as the DM candidate, in which the new gauge boson $Z^\prime$ only interacts with the electron and muon and couples with DM~\footnote{Other studies assuming gauged flavor interactions, see~\cite{gldm-1,gldm-2,gldm-3,Dev:2013hka,gldm-4,hexg,gldm-5,maxim,gldm-6,gldm-7,gldm-8}}. The DM can directly annihilate into leptons through $s$ channel via $Z^\prime$ boson or into a pair of $Z^\prime$ through $t$-channel offering a possible explanation to DAMPE excess in $e^+ + e^-$ data. The structure of this paper is organized as follows. In Section~\ref{section2}, we give a brief introduction of our model. In Section~\ref{section3}, we present the numerical results and discussions. Finally, we draw our conclusions in Section~\ref{section4}.


\section{Model}\label{section2}
In the SM, the difference of electron and muon lepton numbers, $L_e - L_\mu$, can be gauged without gauge anomalies~\cite{model}. In fact one can gauge any combination of $L_i - L_j$ with $i, j = e, \mu, \tau$ without new gauge anomalies produced. The gauge boson $Z^\prime$ resulting from such a gauge symmetry only couples to one of the pairs $e$ and $\mu$ at the tree-level. To have the $Z^\prime$ to couple to dark matter, we introduce a new vector-like fermion $\psi$ as DM with a non-trivial $Y^\prime = L_i - L_j$ number $a$. Demanding that the fermion DM to be vector-like guarantees that our model is gauge anomaly free. The $Z^\prime$ boson can obtain the mass from spontaneous $U(1)_{e-\mu}$ symmetry breaking of a scalar $S$ with a non-trivial charge $Y^\prime = b$. With the new particles $Z^\prime$, $S$ and $\psi$, one can write down the following interactions $L_{new}$,
\begin{eqnarray}
{\cal L}_{new} &=& - {1\over 4} Z^{\prime\mu\nu} Z^\prime_{\mu\nu} + \sum_l \bar l \gamma^\mu (- g^\prime Y_l^\prime  Z^\prime_\mu)l \nonumber \\
&+& \bar \psi [\gamma^\mu (i\partial_\mu - a g^\prime Z^\prime_\mu) - m_\psi]\psi \nonumber\\
&+& (D_\mu S)^\dagger (D^\mu S) + \mu^2_S S^\dagger S + \lambda_S (S^\dagger S)^2 \nonumber \\
&+& \lambda_{SH} (S^\dagger S) H^\dagger H\;,
\end{eqnarray}
where $l$ is summed over the SM leptons. $H$ is the SM Higgs doublet. Then, we can have $Z^\prime$ coupling to fermions given by
\begin{eqnarray}
{\cal L} &=& - g^\prime(a \bar \psi\gamma^\mu \psi + \bar l_i \gamma^\mu l_i - \bar l_j \gamma^\mu l_j
+ \bar \nu_i \gamma^\mu L \nu_i \nonumber \\ &-& \bar \nu_j \gamma^\mu L \nu_j) Z^\prime_\mu\;.
\end{eqnarray}
Since the $Z^\prime$ coupling to leptons are flavor diagonal, there is no tree level flavor changing neutral current induced by $Z^\prime$ in our model.

After symmetry breaking, the physical component of the scalar fields $S$ and $H$ can be written as $(v_S+s)/\sqrt{2}$ and $(v+h)/\sqrt{2}$,
respectively, where $v_S$ and $v$ are non-vanishing vacuum expectation values. The mass of $Z^\prime$ is given by $m^2_{Z^\prime} = b^2 g^{\prime 2} v_S^2$. For simplicity, we assume a small mixing between $S$ and $H$ by decoupling $S$. Therefore, the above Higgs interactions will not affect our following discussions.

\section{Numerical results and discussions}\label{section3}
In Ref.~\cite{yuan2017}, it was pointed out that the excess of $e^+ + e^-$ flux in DAMPE can be interpreted by a 1.5 TeV DM with annihilation branching ratio $e:\mu=1:1$ without conflicting with other cosmic rays and CMB constraints which we require our model to achieve. If DM also annihilate into $\tau$ pairs, its ratio should be much smaller than $e^+ + e^-$ pairs. For this reason, among the models of gauged $L_e - L_\mu$, $L_\mu - L_\tau$ and $L_e -L_\tau$, only $L_e - L_\mu$ will work. This is the model we will study in the following. We implement \textsf{FeynRules}~\cite{feynrule} for detailed model calculations. For the evaluation of DM relic density we use \textsf{MicrOMEGAs}~\cite{micromega}. The results for relevant parameters which producing the required DM relic density are show in Fig. \ref{fig:constraints}.

\begin{figure}[ht]
  \centering
   \includegraphics[width=1.5in]{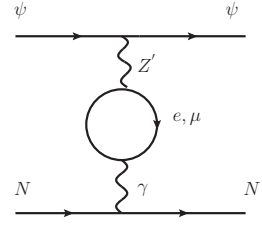}
 \caption{Feynman diagrams for DM $\psi$ scattering with nucleus.}
\label{dd}
\end{figure}
We now consider constraint from direct DM search. Naively, since there is no tree level $Z^\prime$ couplings to quarks, one may expect that there is not much constraint can be obtained from such a consideration. However, $Z^\prime$ couplings to quarks can be generated at one loop level as shown in Fig.~\ref{dd} and lead to DM-nuclus scattering cross section given by,
\begin{eqnarray}
\sigma_{\chi N} &=&\frac{\mu_N^2}{9\pi} (\frac{\alpha_{em}Z}{\pi m_{Z^\prime}^2})^2({\rm log}(\frac{m_e^2}{\mu^2})-{\rm log}(\frac{m_\mu^2}{\mu^2}))^2(ag^\prime g^\prime)^2 \nonumber \\
&=&\frac{\mu_N^2}{9\pi} (\frac{\alpha_{em}Z}{\pi m_{Z^\prime}^2})^2{\rm log}^2(\frac{m_e^2}{m_\mu^2})(a{g^\prime}^2)^2
\label{dm-N-emu}
\end{eqnarray}
where $m_N$ and $Z$ are nucleus's mass and charge, respectively. $\mu_N=\frac{m_\chi m_N}{m_\chi+m_N}$ is the reduced mass of DM-nucleus system. The renormalization scale of this model $\mu$ is set as $\mu=m_{Z^\prime}$. It can be seen that the 1-loop DM-nucleus depends the ratio of $m_e/m_\mu$ rather than renormalization scale. It turns out that the one loop generated DM-nucleus cross section provide strong constraint to allowed $Z^\prime$ mass.

Note that there is a cancellation factor log($m_e/m_\mu$) due to electron and muon contribution in the loop. If such a factor is not there, that is, the coupling of $Z^\prime$ to $e$ and $\mu$ are the same for example, the resulting cross section would be even larger leading to even larger portion of $Z^\prime$ mass being ruled out by direct DM search effect. Combining information on DM relic density and direct DM detection results, we obtain constraints of the relevant parameters which are shown in Fig.~\ref{fig:constraints}.

Besides, the $Z^\prime$ can also induce the process $e^+e^- \to \ell^+\ell^-$, which is strongly constrained by LEP measurements of four-lepton contact interactions \cite{lep2} and di-lepton resonance searches in $e^+e^- \to \ell^+\ell^-\gamma$ \cite{Abbiendi:1999wm}. Following the analysis in Ref.~\cite{Freitas:2014pua}, we can derive the following bounds of the coupling and mass of $Z^\prime$ at 90\% C.L.,
\begin{eqnarray}
g' /m_{Z^\prime} <
\begin{cases}
2.0\times 10^{-4} {\rm GeV}^{-1}, &m_{Z^\prime}>200 {\rm~GeV} \cr 6.9\times 10^{-4} {\rm~GeV}^{-1}, & m_{Z^\prime}\in[100, 200] {\rm~GeV}
\end{cases}
\end{eqnarray}
At a future linear $e^+ e^-$ collider with a CM energy up to 1 TeV, such as ILC, the sensitivity to leptophilic DM is expected to increase significantly with respect to LEP. By re-scaling the LEP limits, one can estimate the ILC bounds at the $90\%$ C.L.~\cite{Freitas:2014pua},
\begin{eqnarray}
g' /m_{Z^\prime} <
\begin{cases}
2.2\times 10^{-5} {\rm GeV}^{-1}, &m_{Z^\prime}>1000 {\rm~GeV} \cr 7.6\times 10^{-5} {\rm~GeV}^{-1}, & m_{Z^\prime}\in[100, 1000] {\rm~GeV}
\end{cases}
\end{eqnarray}

\begin{figure}[h]
  \centering
   \includegraphics[width=3in,height=2in]{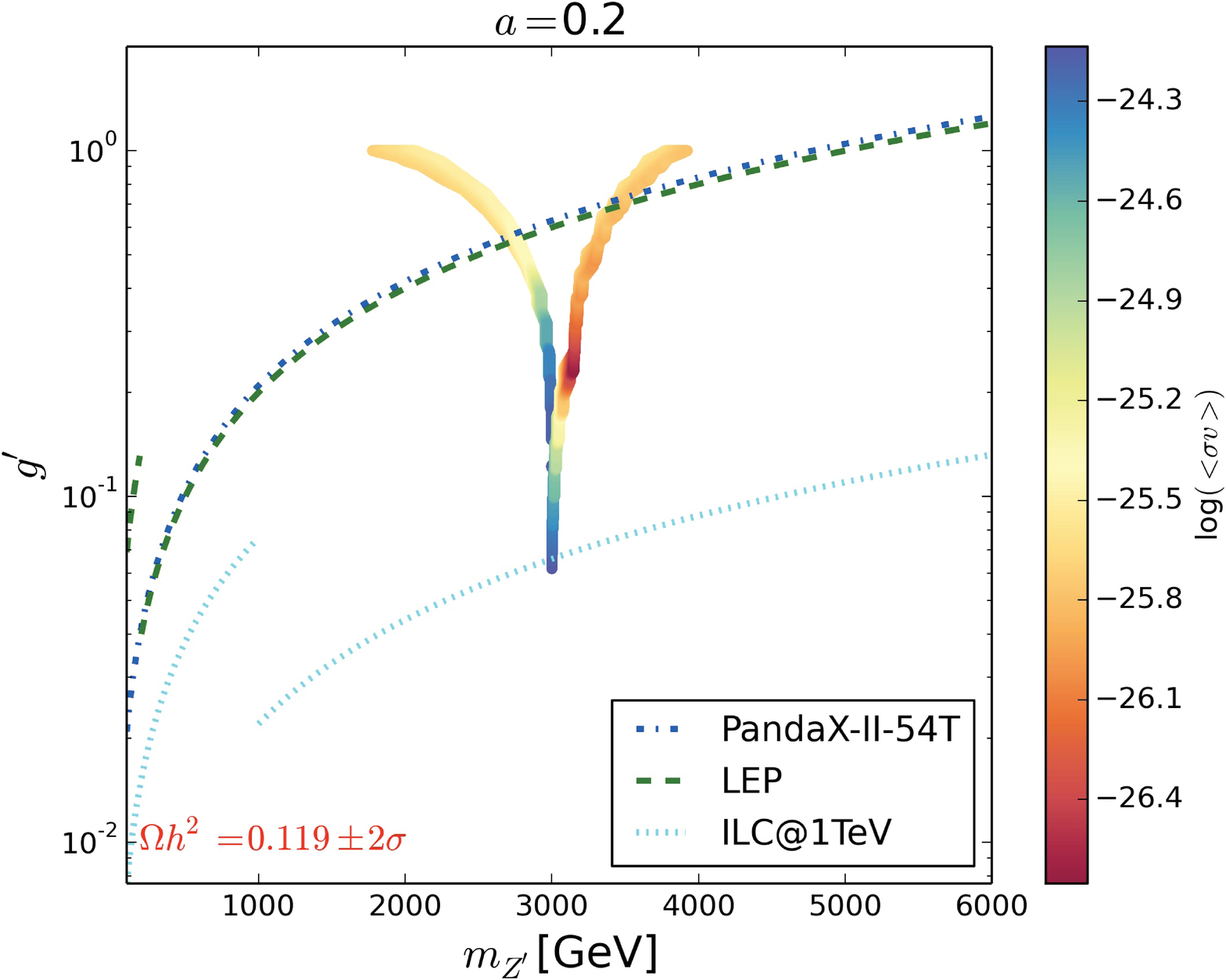}
   \includegraphics[width=3in,height=2in]{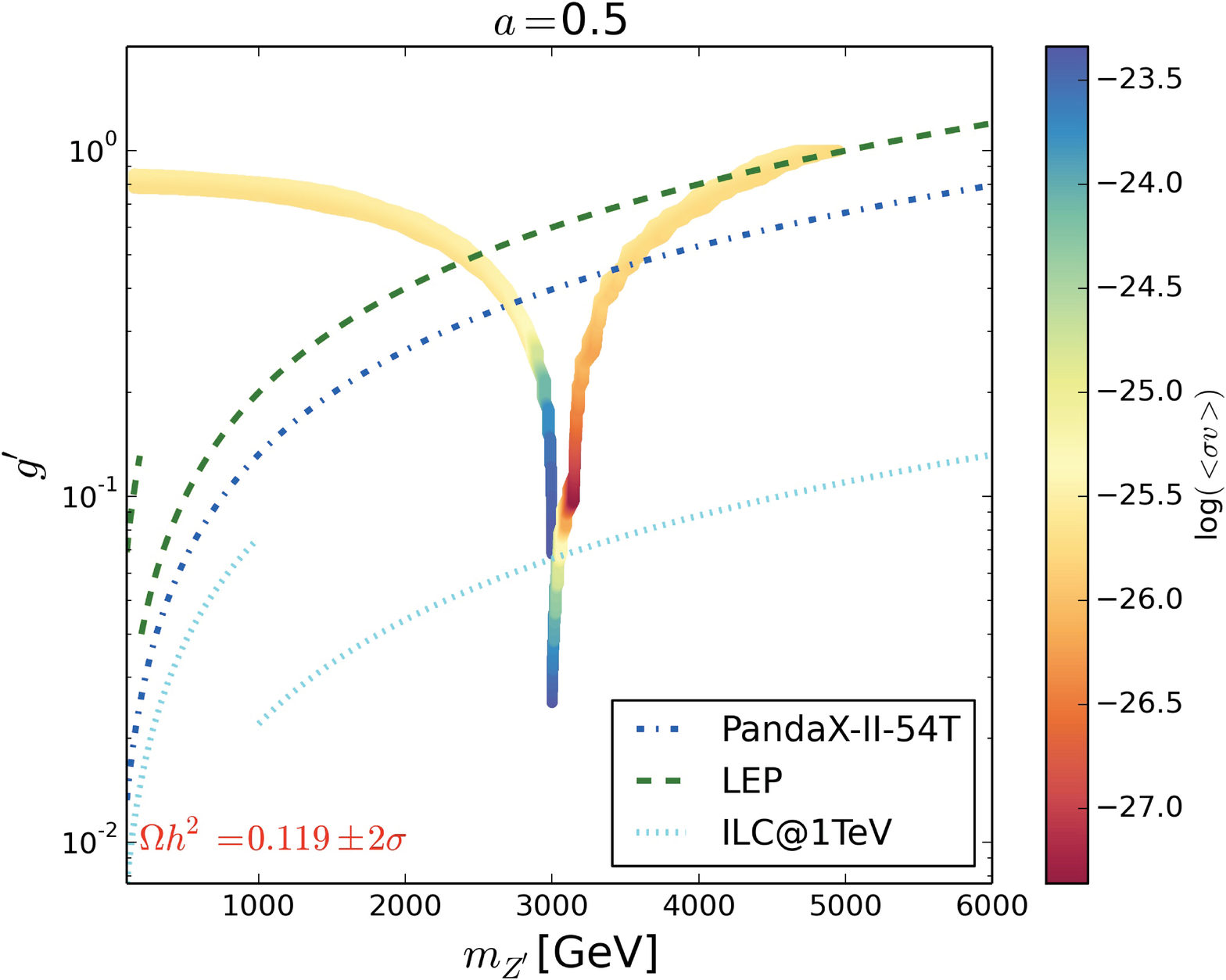}
   \includegraphics[width=3in,height=2in]{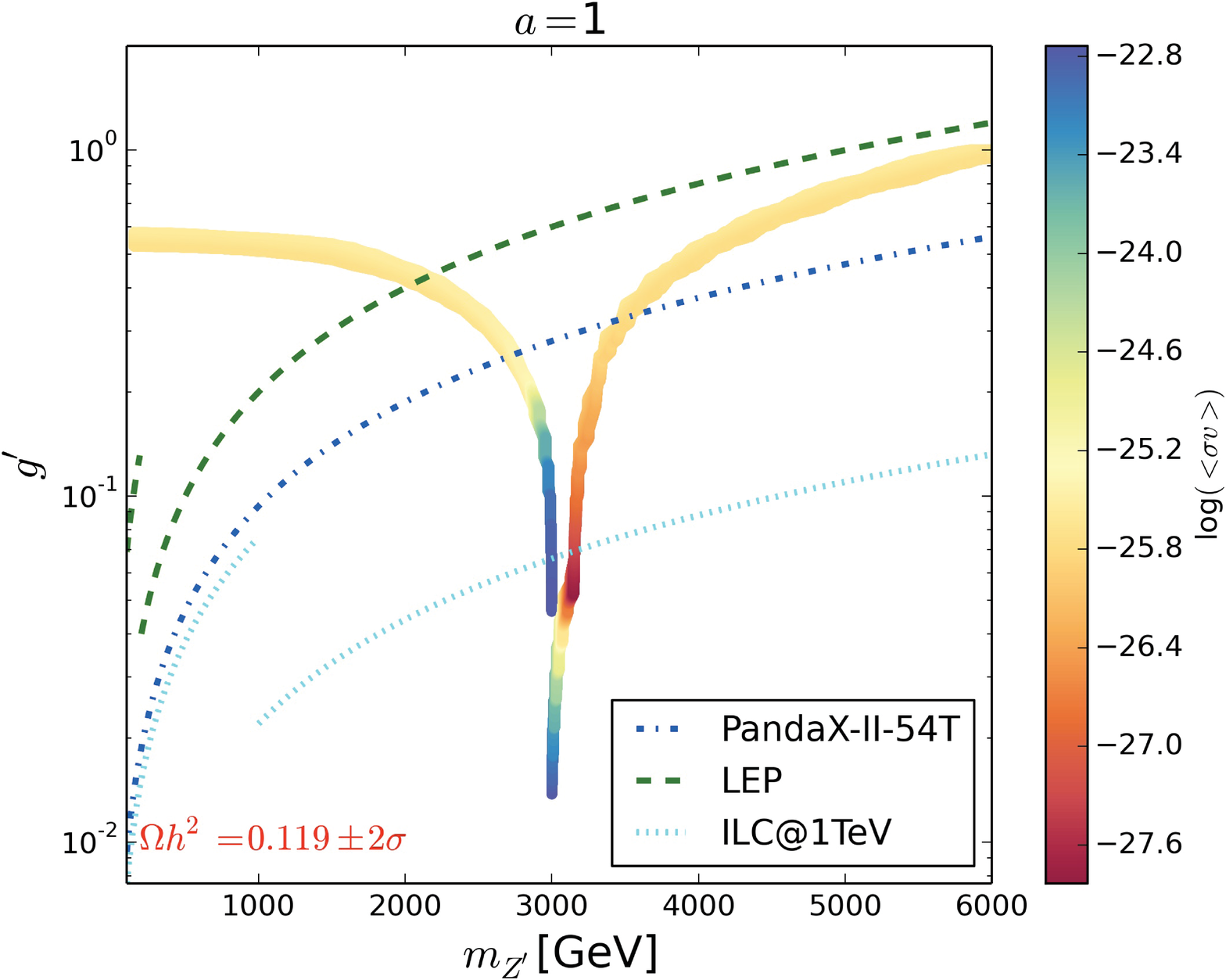}
\vspace{-0.3cm}
\caption{Constraints of DM relic density, LEP, ILC-1 TeV and direct detection. The bands satisfy the DM relic density within the 2$\sigma$ range~\cite{planck}. The colormap denotes the current annihilation cross section $\langle \sigma v \rangle$. The regions above the blue dashed-dotted, green dashed and cyan dotted curves can be excluded by 90\% C.L. exclusion limit from PandaX-II~\cite{Pandax}, LEP data~\cite{lep2} and ILC-1 TeV, respectively.}
\label{fig:constraints}
\end{figure}

In Fig.~\ref{fig:constraints}, we plot the bands that can produce the DM relic density within $2\sigma$ ranges on the planes of the gauge coupling $g^\prime$ versus $Z^\prime$ mass. The blue dash-dotted and green dashed curves are 90\% C.L. upper limits of PandaX-II and LEP for $a=1,0.5,0.2$. We can see that the PandaX-II data have imposed strong constraints on the $Z^\prime$ mass, which is stronger than the LEP bound for $a>0.2$. The allowed mass ranges of $m_{Z^\prime}$ is about $3000 \pm 500$ GeV and the coupling $g^\prime<0.5$ for $0.2 <a<1$. The corresponding DM annihilation cross sections $\langle \sigma v \rangle$ in the present Universe vary from $O(10^{-26}) \sim O(10^{-24})~cm^3/s$. When $m_{Z^\prime} \simeq 3300$ GeV, the DM annihilation cross section can reach $3 \times 10^{-26}~cm^3/s$, which is typical thermal DM annihilation cross section and is assumed in fitting DAMEP excess in~\cite{yuan2017}. Such favored regions will be further tested by measuring four-lepton contact interactions in the future ILC experiment.

\begin{figure}[ht]
  \centering
   \includegraphics[width=3in]{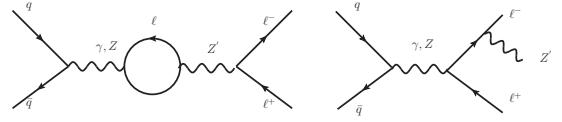}
 \caption{Feynman diagrams for the Drell-Yan process induced by $Z^\prime$ (left panel) and associated production process $Z^\prime\ell^+\ell^-$ (right panel) at the LHC.}
\label{feyn-dy}
\end{figure}

\begin{table}[ht!]
\small
\begin{tabular}{|c|c|c||c|c|c|c|c|}
  \hline
  $m_{Z^\prime}$ (TeV)& $a$ &$g^\prime$ & $\Delta a_\mu \times 10^{15}$ &$\sigma_{l^+ l^-}(fb)$ &$\sigma_{ Z^\prime l^+ l^-}(fb)$ \\

\hline\hline

   $3.15$ &0.5 & 0.10 & 95 &$1.48\times10^{-6}$&$3.40\times10^{-9}$\\
   $3.00$ &1.0 & 0.015 & 2.3 &$4.46\times10^{-8}$&$1.24\times10^{-10}$\\
   $3.15$ &0.2 &0.26 & 647 &$1.01\times10^{-5}$&$2.32\times10^{-8}$\\
   \hline
\end{tabular}
\caption{The cross sections of Drell-Yan process $pp \to Z^\prime \to \ell^+\ell^-$ and the associated production $pp \to Z^\prime\ell^+\ell^-$ at 13 TeV LHC. The corrections to muon $g-2$ ($\Delta a_{\mu}$) are also given. The benchmark points satisfy the DM relic density, the LEP bound, the DAMPE $e^+ + e^-$ flux excess and the PandaX limit.}
\label{tab}
\end{table}

We now study the possible collider signatures  of this model. $Z^\prime$ be produced at the LHC and can induce the Drell-Yan process because of the loop-induced couplings between the mediator and light quarks~\cite{DEramo:2017zqw} shown in Fig.~\ref{feyn-dy} (left panel). 
The cross section in the narrow width limit is given by,
\begin{eqnarray}
\sigma_{p p \to Z^\prime \to l^+ l^-} = \frac{\pi \, BR_{Z^\prime \to l^+ l^-}}{3 s}
\sum_{q} C_{q\bar{q}} (m_{Z^\prime}^2 / s) \, {g^{V}_{q}}^2,
\end{eqnarray}
with
\begin{eqnarray}
g^{V}_{q}=\frac{\alpha_{em}}{3\pi}Q_qg^\prime {\rm log}(\frac{m_e^2}{m_\mu^2}),
\label{eq:sxLHCschannelNarrow}
\end{eqnarray}
where $BR_{Z^\prime \to l^+ l^-}$ is the branching ratio of the decay $Z^\prime \to l^+ l^-$ and $s$ is the squared colliding energy. $Q_q$ is the electric charge of quarks. The parton luminosity $C_{q\bar{q}} (m_{Z^\prime}^2 / s)$ for the quark $q$ reads
\begin{eqnarray}
C_{q\bar{q}}(y) = \int_{y}^1 d x \, \frac{f_{q}(x) \, f_{\bar{q}}(y / x) + f_{q}(y / x) \, f_{\bar{q}}(x)}{x}  \ ,
\end{eqnarray}
with $f_{q, \bar{q}}(x)$ the quark and antiquark parton distribution function (PDF). We utilize MRST~\cite{mrst} to calculate the PDFs. We calculate the loop induced process $pp \to Z^\prime \to \ell^+\ell^-$ for several surviving samples at 13 TeV LHC, as shown in Tab~\ref{tab}. It can be seen that these cross sections are much lower than the LHC-13 TeV sensitivity~\cite{Aaboud:2017buh} due the cancellation between electron loop and muon loop.

In Tab~\ref{tab}, we also show the results of associated production process $pp \to Z^\prime \ell^+\ell^-$ (induce by the right panel in Fig. 2), which can give four leptons or two leptons with large missing transverse energy signatures at the LHC. However, the cross sections are negligible small at 13 TeV LHC.

Exchange of  $Z^\prime$ can also induce a non-zero contribution to muon $g-2$, which is given by
\begin{eqnarray}
\Delta a_\mu=\frac{g-2}{2} = \frac{{g^\prime}^2}{12\pi^2} \frac{m_\mu^2}{m_{Z^\prime}^2} ,
\end{eqnarray}
The values of $\Delta a_{\mu}$ is shown in Tab~\ref{tab}. We see that the $Z^\prime$ contribution is less than the value required by explaining the deviation of the muon $g-2$ from its experimental measurement.

\section{Conclusions}\label{section4}
In this paper, we studied recent DAMPE cosmic-ray eletron+positron excess in the gauged $L_e-L_\mu$ model with Dirac ferimon DM. Our $U(1)_{e-\mu}$ gauge boson $Z^\prime$ connects the DM with the SM leptons and can accommodate the DAMPE excess. The direct DM detection appearing at one loop level can rule out a large portion of parameter space that satisfying DM relic density and DAMPE data, although there is cancellation between $Z^\prime ee$ and $Z^\prime \mu\mu$ couplings. We found that our model can fit the DAMPE data without conflicting with the current direct detection limits. We also discussed the LHC signatures, including $pp \to Z^\prime \to 2\ell$ and $pp \to 2\ell + Z^\prime$ and the muon $g-2$ anomaly and find that the $Z^\prime$ effects are small.

\section{acknowledgments}
G. Duan was supported by a visitor program of Nanjing Normal University, during which this work
was finished. This work was supported by the National Natural Science Foundation of China (NNSFC)
under grant No. 11705093 and 11675242, by the CAS Center for Excellence in Particle
Physics (CCEPP),  by the CAS Key Research Program of Frontier Sciences and by a Key R\&D Program
of Ministry of Science and Technology of China under number 2017YFA0402200-04.
XG was supported in part by Key Laboratory for Particle Physics,
Astrophysics and Cosmology, Ministry of Education, and Shanghai Key Laboratory for Particle
Physics and Cosmology (Grant No. 15DZ2272100), in part by the NSFC (Grant Nos. 11575111 and 11735010) and also in part by the MOST (Grant No. MOST104-2112-M-002-015-MY3).

\end{document}